\documentclass[a4paper,11pt]{article}
\usepackage{pos}
\usepackage[percent]{overpic}

\title{$K^*/K$ ratio and the time between freeze-outs for intermediate-mass Ar+Sc system at the SPS energy range}

\author*[a]{Bartosz Kozłowski for the NA61/SHINE Collaboration}

\affiliation[a]{Warsaw University of Technology, Faculty of Physics,\\
  Koszykowa 75, 00-662 Warsaw, Poland}

\emailAdd{bartosz.kozlowski@cern.ch}

\abstract{
Resonance production is one of the key observables to study the dynamics of high-energy collisions. In dense systems created in heavy nucleus-nucleus collisions, the properties of some of them (widths, masses, branching ratios) were predicted to be modified due to partial restoration of chiral symmetry. The resonance spectra and yields are also important inputs for Blast-Wave and Hadron Resonance Gas models. Finally, the analysis of strange $K^*(892)^0$ resonance allows us to better understand the time evolution of high-energy nucleus-nucleus collision. Namely, the ratio of $K^*(892)^0$ to charged kaons is used to determine the time between chemical and kinetic freeze-outs.

In this article, the first results on the analysis of $K^*(892)^0$ production in central Ar+Sc collisions at three SPS energies ($\sqrt{s_{\mathrm{NN}}}$ = 8.8, 11.9, 16.8 GeV) are presented. The $K^*(892)^0/K^{+/-}$ yield ratios are compared with corresponding results in $p$+$p$ collisions, allowing us to estimate the time between chemical and thermal freeze-outs in Ar+Sc collisions. These first results for intermediate-mass nucleus-nucleus systems at the SPS energy range are compared with the results of heavier systems at a similar energy range.
}

\FullConference{42nd International Conference on High Energy Physics (ICHEP2024)\\
18-24 July 2024\\
Prague, Czech Republic\\}

\begin{document}
\maketitle

\section{Introduction}
NA61/SHINE~\cite{NA61:2014lfx} is a multipurpose, fixed-target experiment located at the SPS (Super Proton Synchrotron) accelerator at CERN. The NA61/SHINE research program covers strong interaction physics and measurements for neutrino and cosmic-ray physics. The main goal of the strong interaction program is to study the properties of the onset of deconfinement and search for the critical point.

Resonance production is a useful tool for studying the dynamics of high-energy collisions, especially the time between freeze-outs~\cite{Markert:2002rw}. The mean lifetime of $K^*(892)^0$ resonance is comparable with the expected time between chemical and kinetic freeze-outs. Thus, some of the $K^*(892)^0$ resonances will decay inside the fireball. The momenta of their decay products can be modified due to elastic scattering, which prevents the reconstruction of $K^*(892)^0$ via the invariant mass analysis. Consequently, the suppression of the observed $K^*(892)^0$ yield is expected.

\section{Transverse momentum/mass spectra and inverse slope parameter}
\begin{figure}[!ht]
    \centering
        \includegraphics[width=0.46\linewidth]{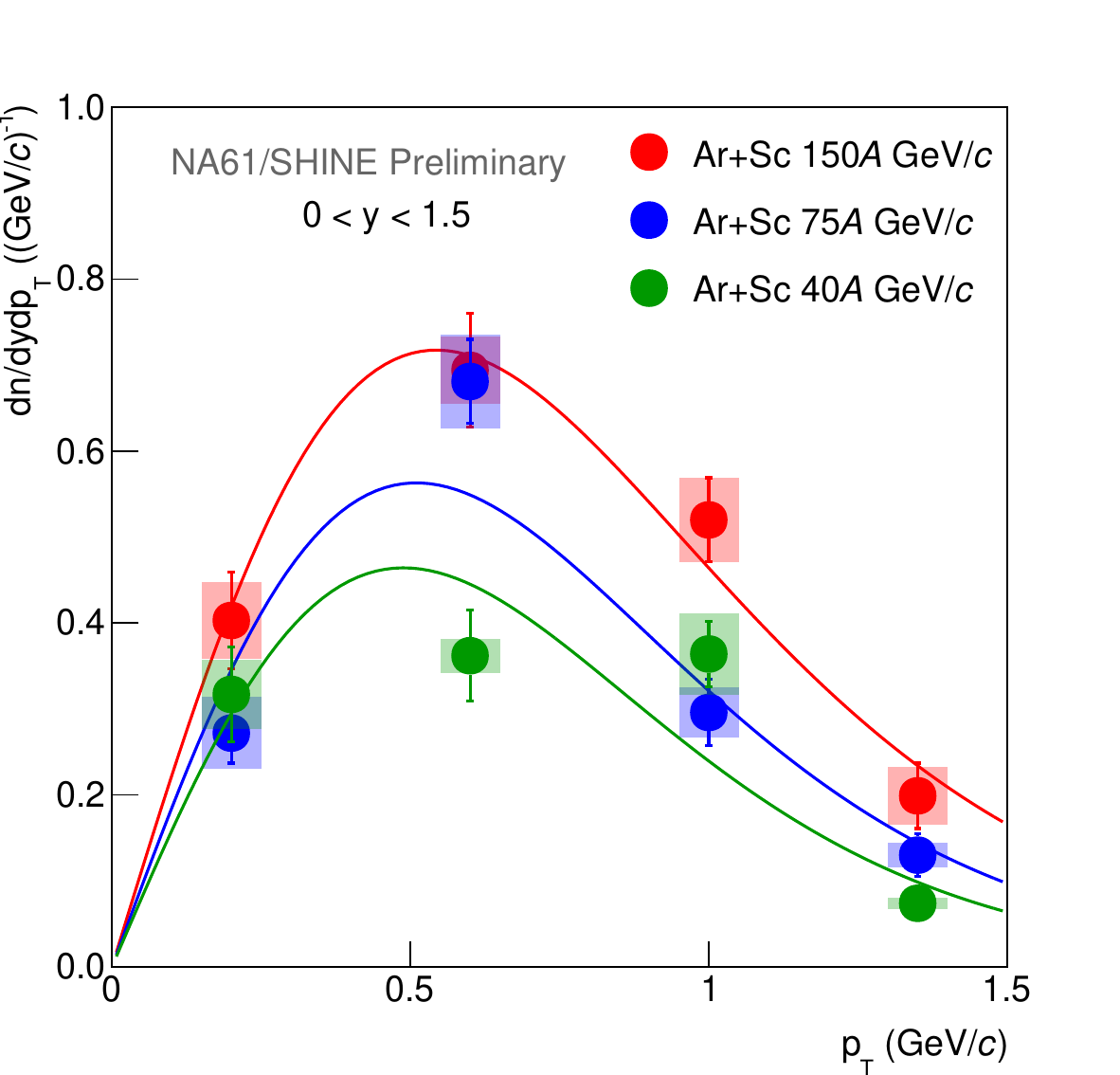}
        \includegraphics[width=0.46\linewidth]{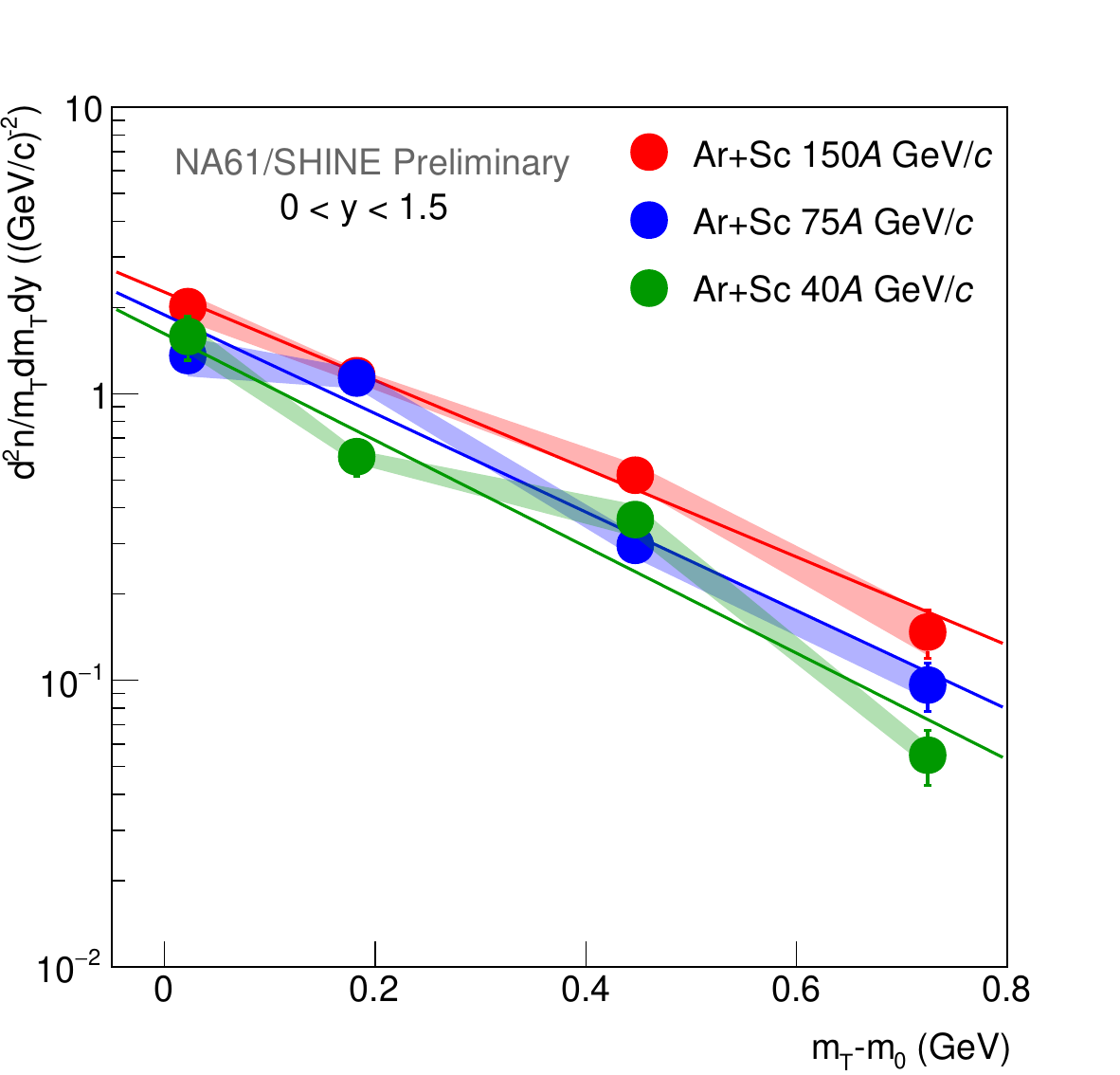}
    \caption{Transverse momentum (left) and transverse mass (right) spectra of $K^*(892)^0$ resonances produced in 0--10\% central Ar+Sc collisions at beam momenta 40$A$, 75$A$, and 150$A$ GeV/$c$ in rapidity range 0 < $y$ < 1.5. Statistical and systematic uncertainties are presented as color vertical bars and boxes/bands, respectively. The color curves represent exponential fits.}
    \label{fig:transverse}
\end{figure}

\begin{figure}[!ht]
    \centering
        \includegraphics[width=0.46\linewidth]{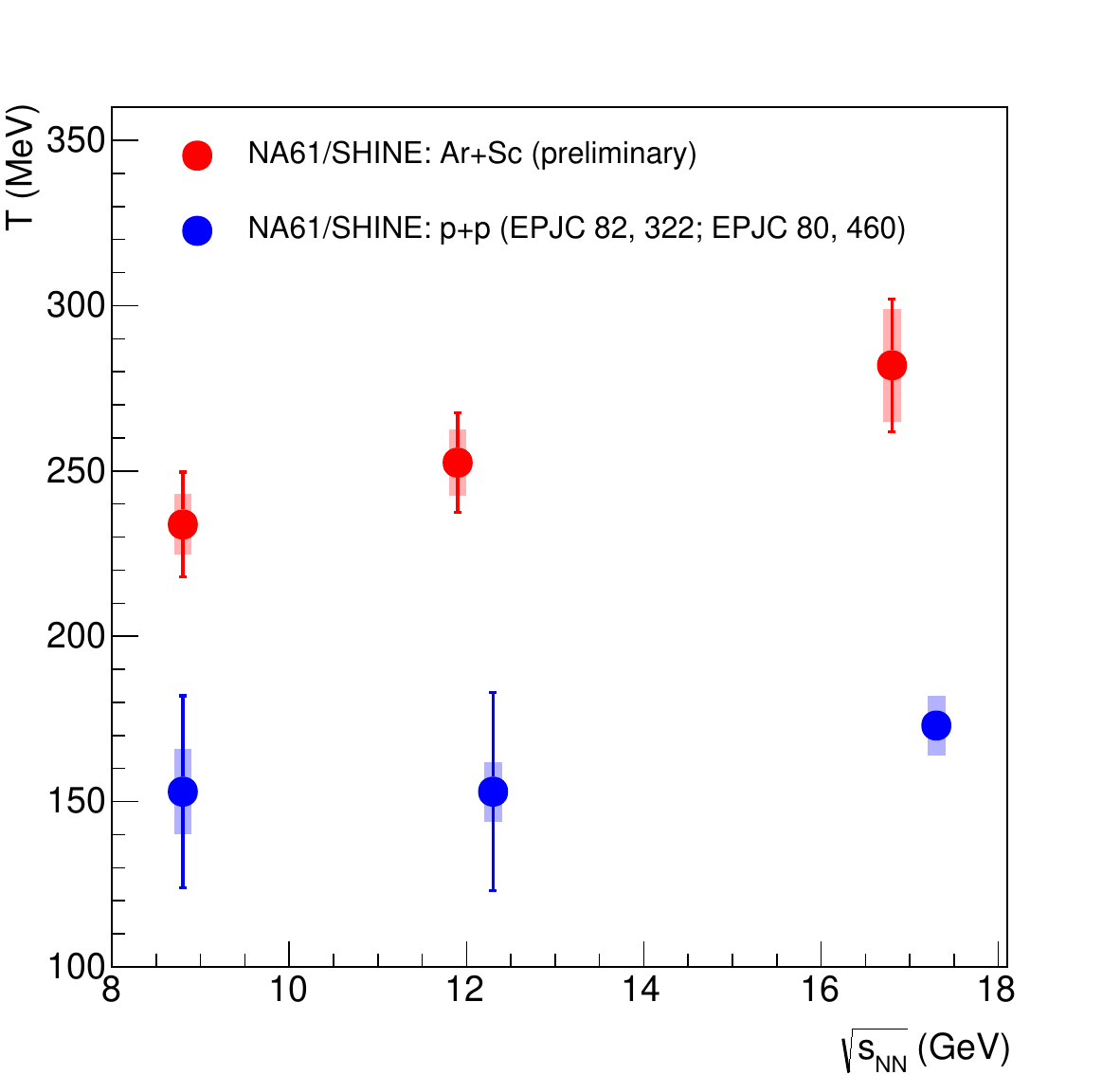}
        \begin{overpic}[width=0.46\linewidth]{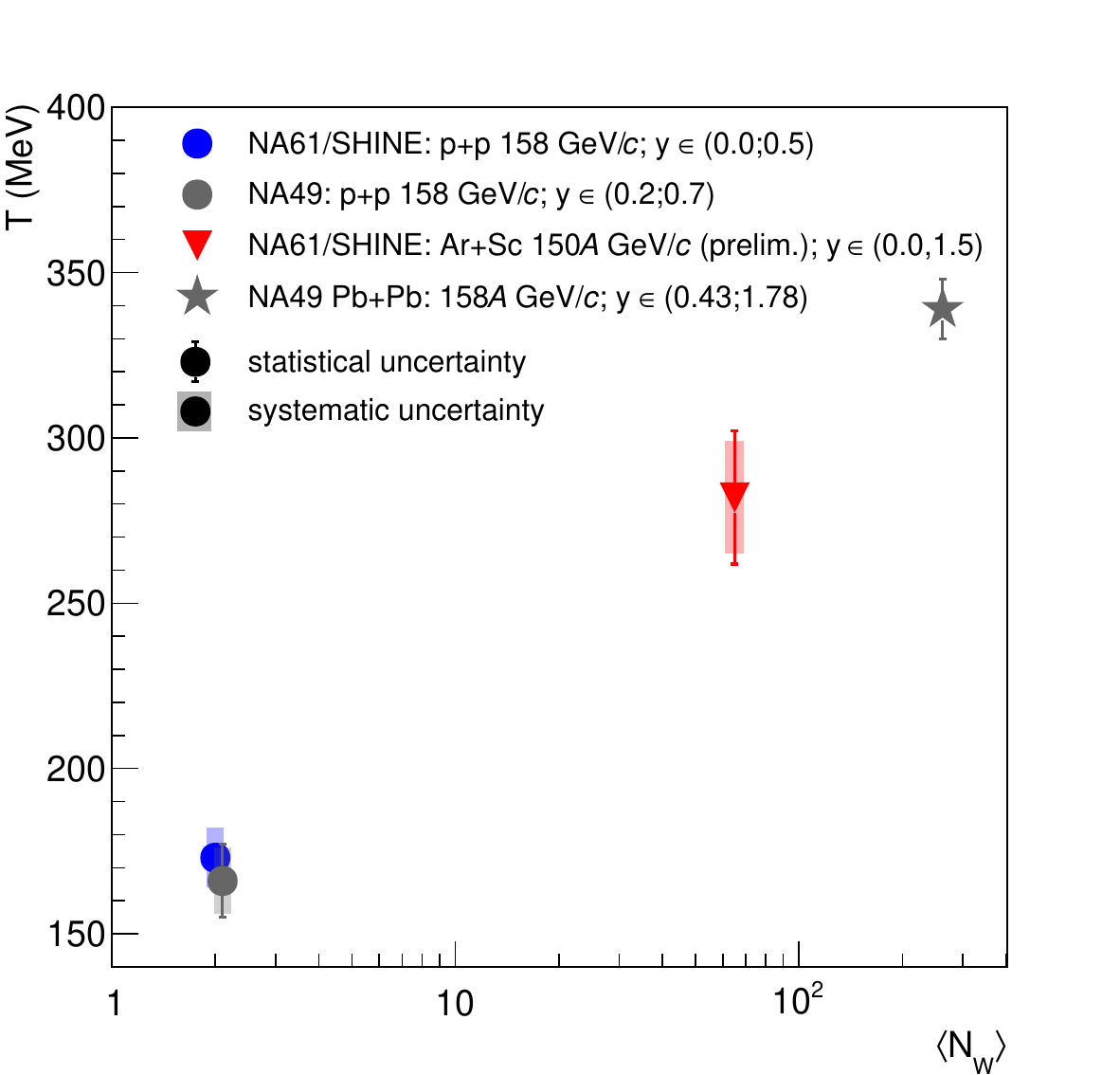}
            \put (34,18) {\tiny $N_{W}$ -- number of wounded nucleons }
        \end{overpic}
    \caption{Inverse slope parameters of $K^*(892)^0$ transverse momentum/mass spectra as a function of collision energy (left) and system size (right). Statistical and systematic uncertainties are presented as color vertical bars and boxes, respectively. }
    \label{fig:transverse:comparison}
\end{figure}

Figure \ref{fig:transverse} shows transverse momentum and transverse mass spectra of $K^*(892)^0$ mesons produced in 0--10\% central Ar+Sc collisions at beam momenta 40$A$, 75$A$, and 150$A$ GeV/$c$ (collision center-of-mass energies $\sqrt{s_{\mathrm{NN}}}$ = 8.8, 11.9, 16.8 GeV). The results were obtained in the center-of-mass rapidity range 0 < $y$ < 1.5. The solid color lines represent fits of exponential functions with inverse slope parameters $T$ presented in Fig.~\ref{fig:transverse:comparison}.

The left panel of Fig.~\ref{fig:transverse:comparison} shows a comparison of the inverse slope parameters obtained from Ar+Sc and $p$+$p$ \cite{NA61SHINE:2021wba,NA61SHINE:2020czr} collisions as a function of collision energy. In the right panel of Fig.~\ref{fig:transverse:comparison} the Ar+Sc result from the highest beam momentum is compared to the results from different collision systems ($p$+$p$ and Pb+Pb) at similar energy \cite{NA61SHINE:2020czr,NA49:2011bfu}. The visible increase of the inverse slope parameter when going from $p$+$p$ to nucleus-nucleus collisions can be explained as due to radial flow.

\section{Rapidity spectra and mean multiplicities}
\begin{figure}[!ht]
    \centering
        \includegraphics[width=0.325\linewidth]{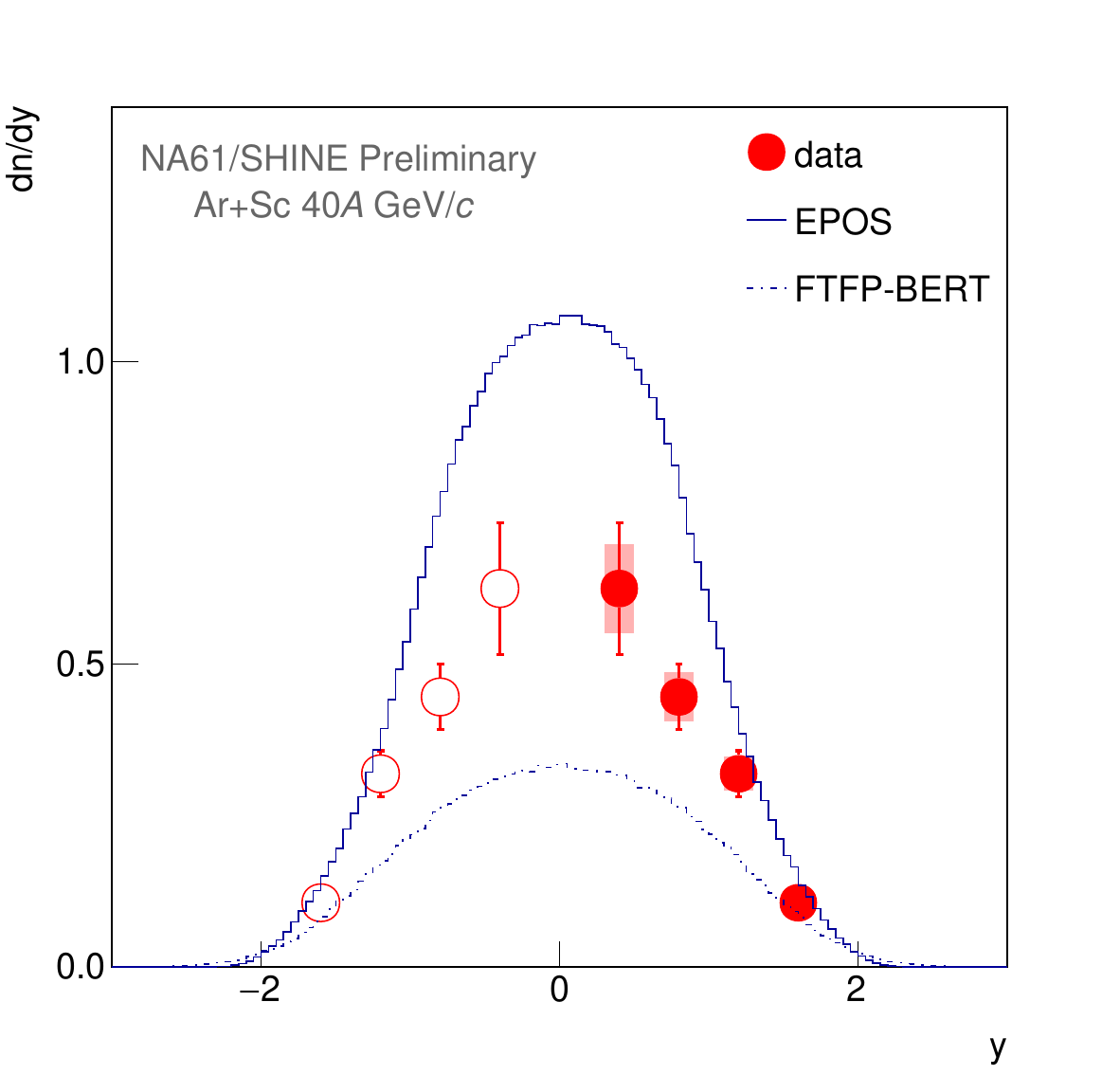}
        \includegraphics[width=0.325\linewidth]{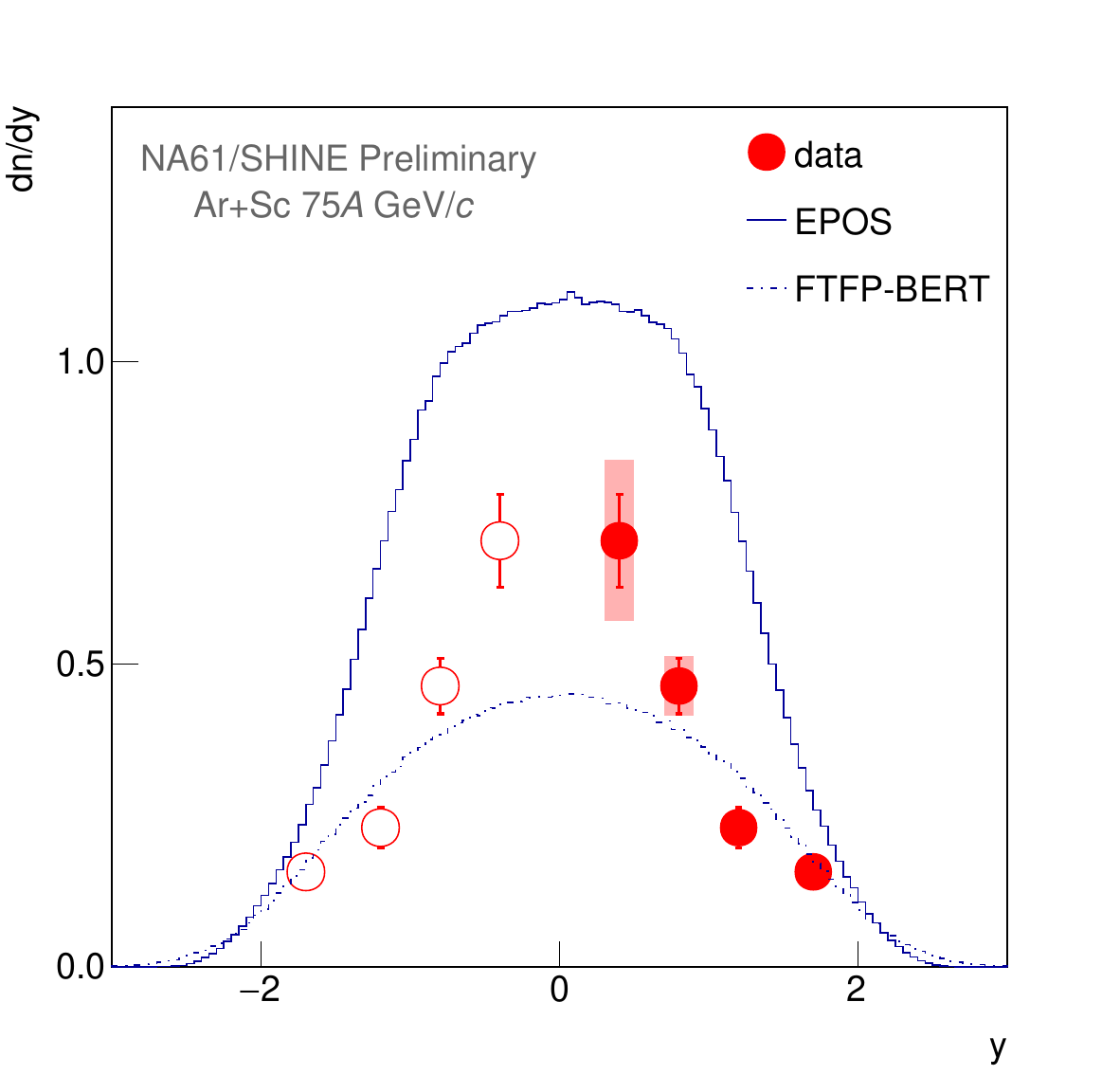}
        \includegraphics[width=0.325\linewidth]{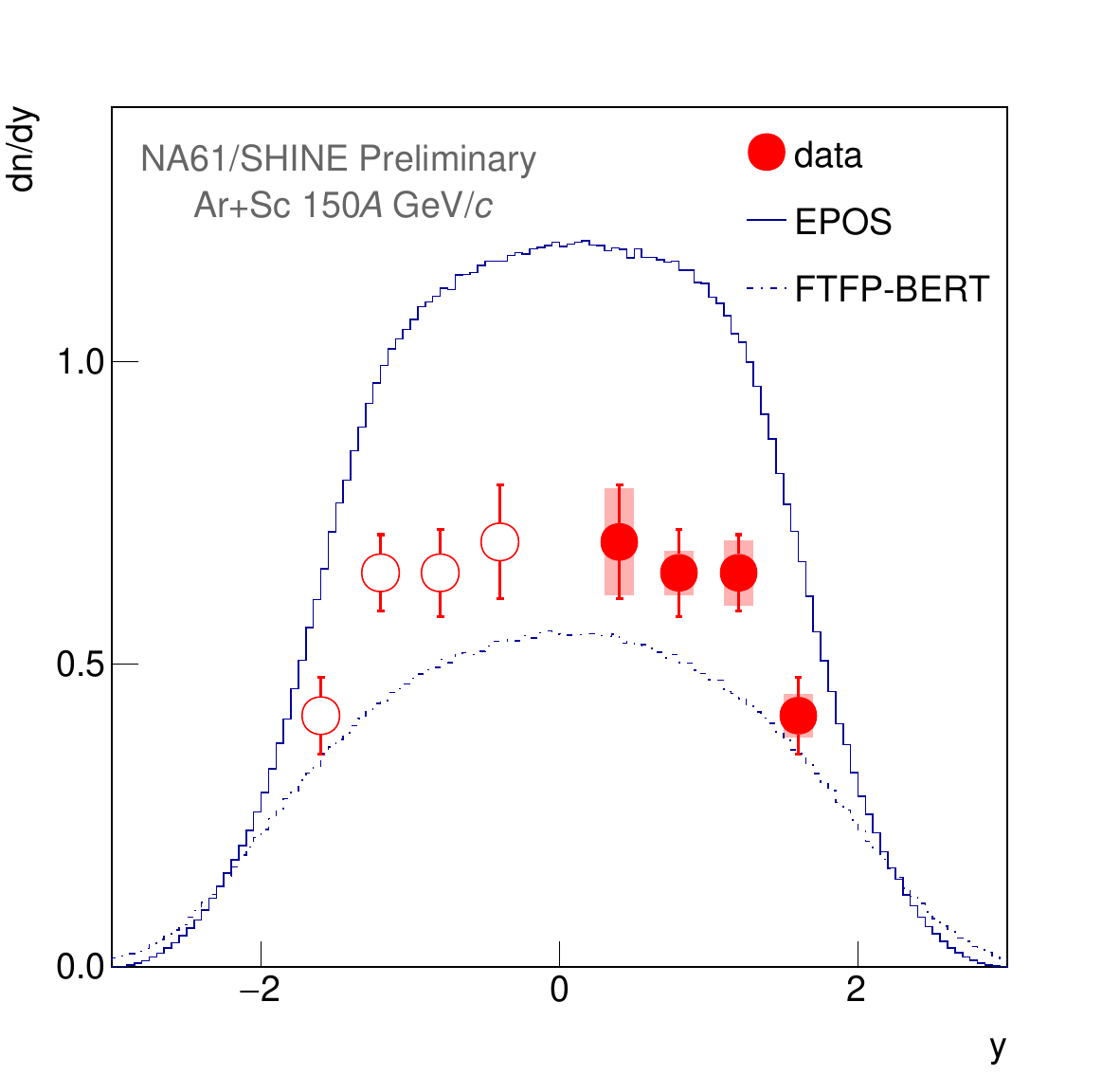}
    \caption{Rapidity spectra of $K^*(892)^0$ resonances produced in 0--10\% central Ar+Sc collisions at beam momenta 40$A$, 75$A$, and 150$A$ GeV/$c$ in transverse momentum range 0 < $p_\mathrm{T}$ < 1.5 GeV/$c$. Statistical and systematic uncertainties are presented as red vertical bars and boxes, respectively. Empty points are reflections around mid-rapidity. Solid and dashed lines represent model predictions.}
    \label{fig:multiplicity}
\end{figure}

Figure~\ref{fig:multiplicity} shows rapidity spectra of $K^*(892)^0$ mesons measured in central Ar+Sc collisions at all three studied beam momenta. The results were obtained in the transverse momentum range 0 < $p_\mathrm{T}$ < 1.5 GeV/$c$. The NA61/SHINE points are compared with predictions of the FTFP-BERT \cite{Allison:2016lfl} and EPOS 1.99 \cite{Pierog:2009zt} models. Both models fail to describe the measured rapidity spectra.

\begin{figure}[!ht]
    \centering
        \includegraphics[width=0.46\linewidth]{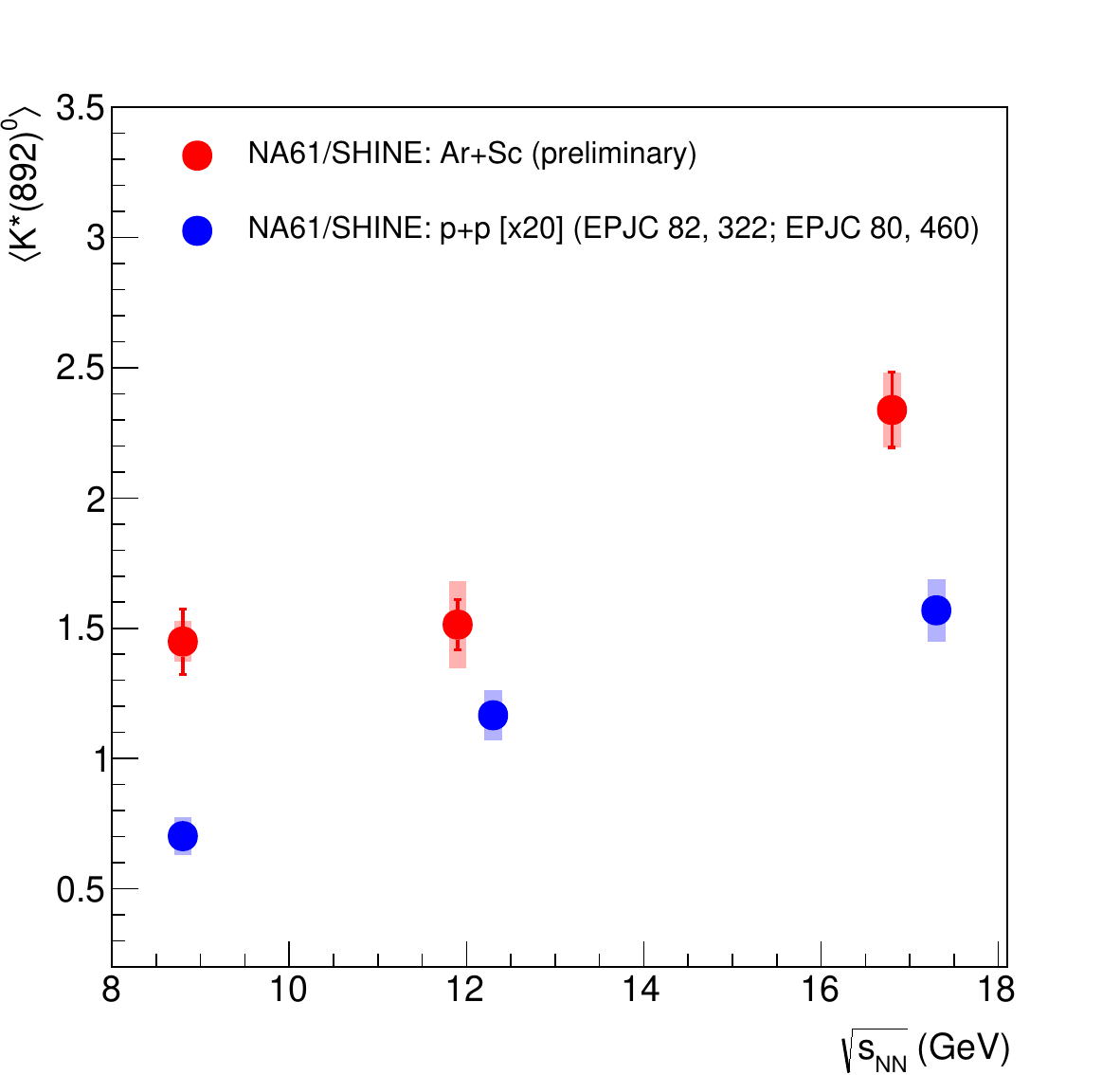}
        \includegraphics[width=0.46\linewidth]{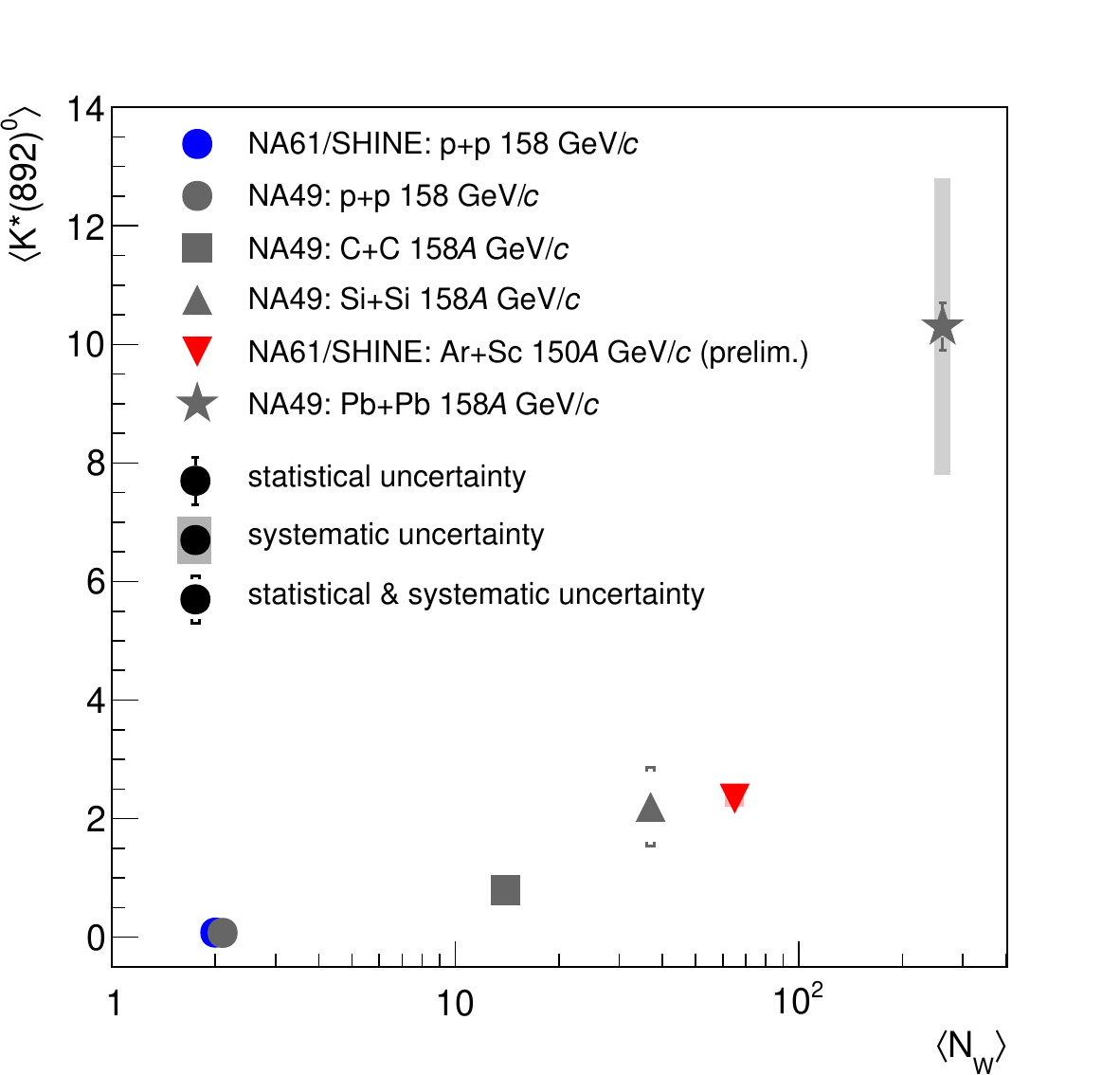}
    \caption{Mean multiplicity of $K^*(892)^0$ resonances as a function of collision energy (left) and system size (right). Statistical and systematic uncertainties are presented as color vertical bars and boxes, respectively.}
    \label{fig:multiplicity:comparison}
\end{figure}

The mean multiplicity of $K^*(892)^0$ mesons was calculated as the sum of measured rapidity points (multiplied by bin widths) scaled under the assumption that the ratio between
measured and unmeasured regions is the same in data and EPOS 1.99.
Figure~\ref{fig:multiplicity:comparison} shows a comparison of mean multiplicities of $K^*(892)^0$ mesons in different collision systems and at different energies. The left panel of Fig.~\ref{fig:multiplicity:comparison} presents a comparison of mean multiplicity of $K^*(892)^0$ obtained in Ar+Sc and $p$+$p$ \cite{NA61SHINE:2021wba,NA61SHINE:2020czr} collisions ($p$+$p$ data are scaled by a factor of 20). The right panel of Fig.~\ref{fig:multiplicity:comparison} shows the multiplicities of $K^*(892)^0$ mesons as a function of the number of wounded nucleons for 150/158$A$ GeV/$c$ beam momentum. The new NA61/SHINE Ar+Sc point is compared to $p$+$p$~\cite{NA61SHINE:2020czr, NA49:2011bfu} and nucleus-nucleus \cite{NA49:2011bfu} data. Both plots show an expected increase of multiplicity with collision energy and system size.

\section{\texorpdfstring{$K^*/K$}{Lg} ratio and the time between freeze-outs}
\begin{figure}[!ht]
    \centering
        \includegraphics[width=0.325\linewidth]{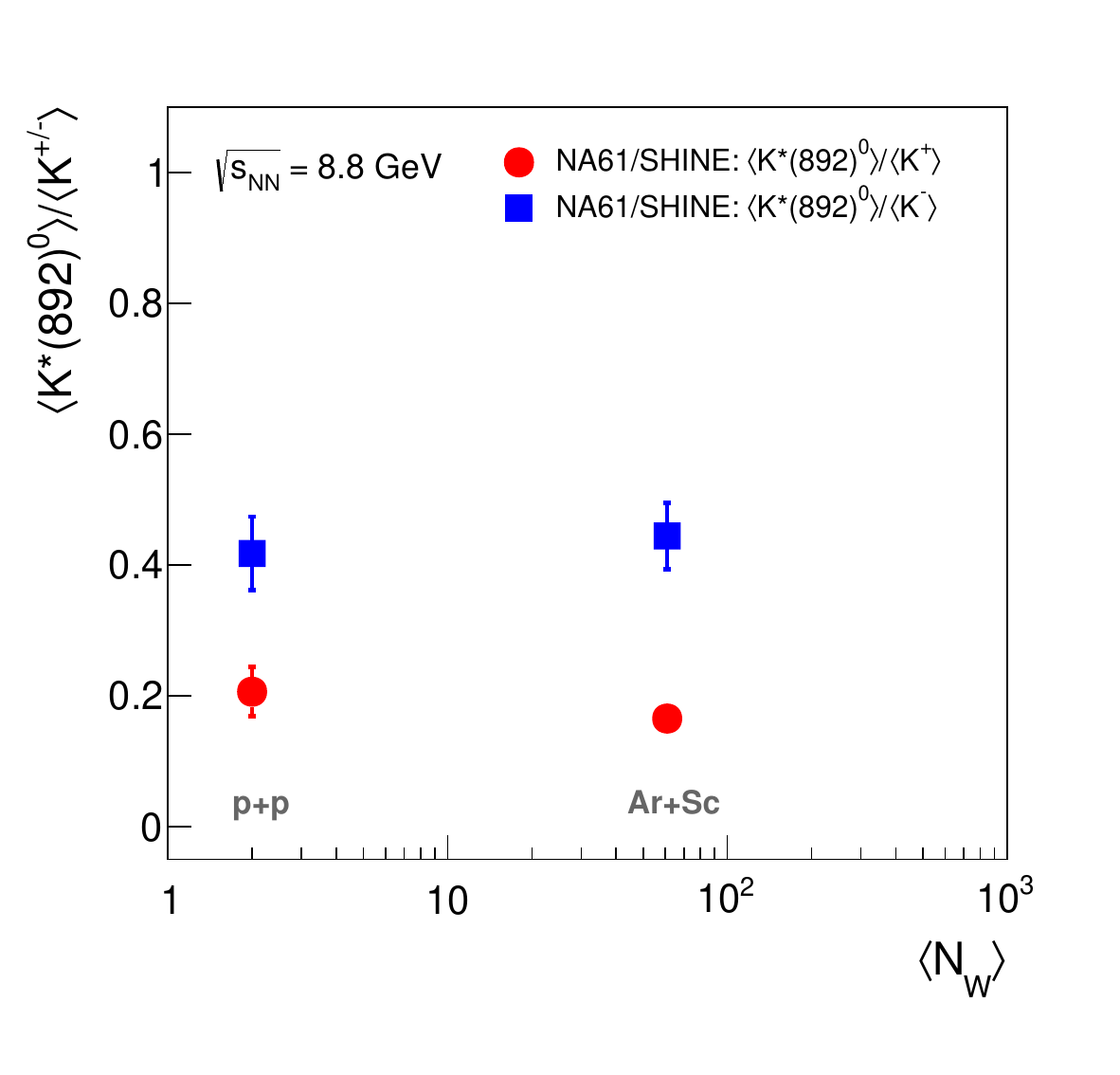}
        \includegraphics[width=0.325\linewidth]{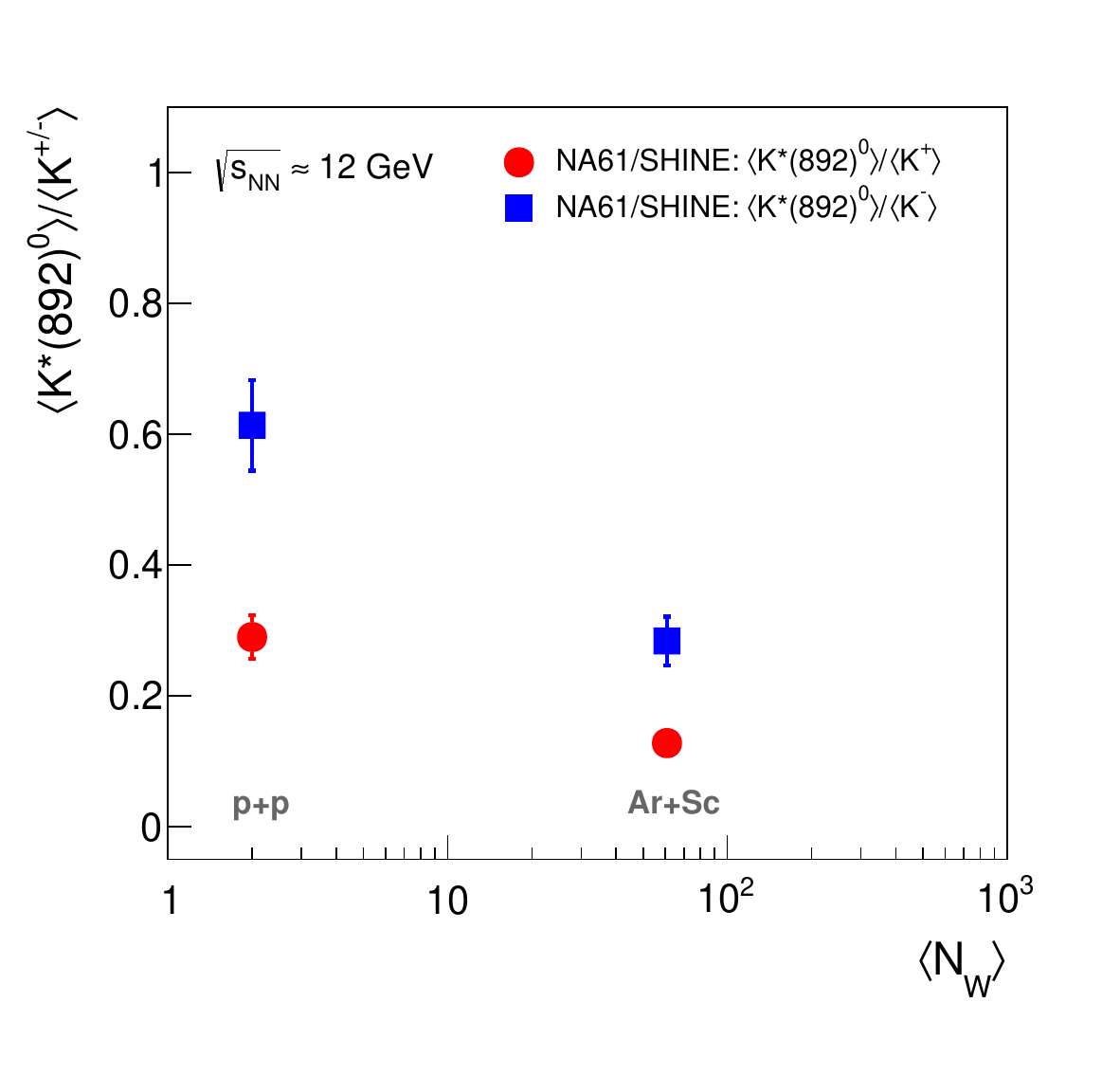}
        \includegraphics[width=0.325\linewidth]{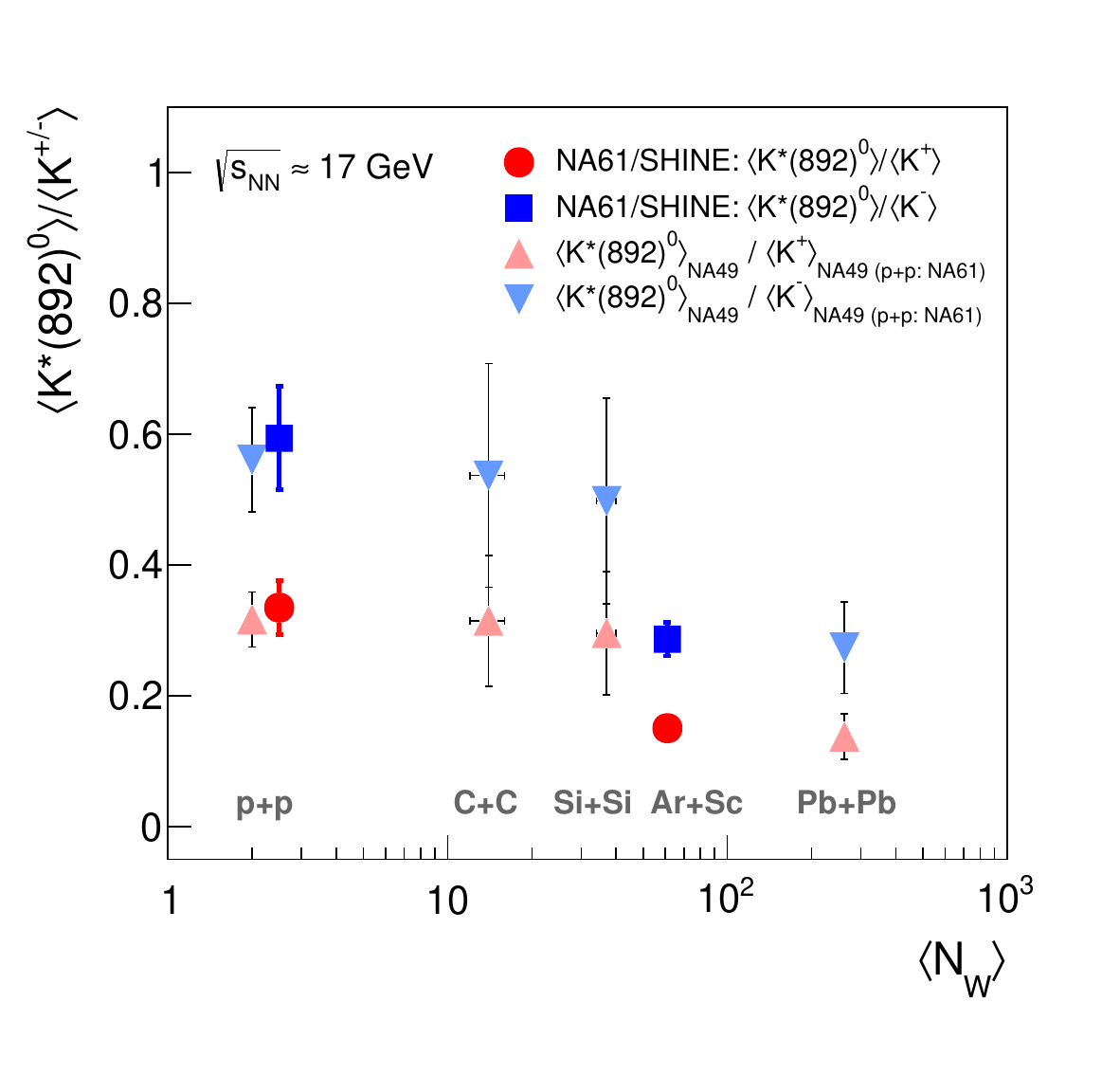}
 \vspace{-0.3cm}
    \caption{$K^*/K$ ratio as a function of the number of wounded nucleons for collision energies 
    $\sqrt{s_\mathrm{NN}}$ = 8.8 GeV, $\sqrt{s_\mathrm{NN}}$ $\approx$ 12 GeV, and $\sqrt{s_\mathrm{NN}}$ $\approx$ 17 GeV. Vertical bars represent total uncertainties.}
    \label{fig:ratio}
\end{figure}

Figure \ref{fig:ratio} shows the system size dependence of $K^*/K$ ratio at collision energies $\sqrt{s_\mathrm{NN}}$ = 8.8 GeV, $\sqrt{s_\mathrm{NN}}$ $\approx$ 12 GeV, and $\sqrt{s_\mathrm{NN}}$ $\approx$ 17 GeV. 
The plots were prepared based on preliminary NA61/SHINE results on $K^{*}(892)^0$ production in 0--10\% central Ar+Sc collisions as well as published NA49 and NA61/SHINE results on $K^{*}(892)^0$~\cite{NA61SHINE:2021wba, NA61SHINE:2020czr, NA49:2011bfu} and $K^{+/-}$~\cite{NA61SHINE:2017fne, NA49:2004jzr, NA61SHINE:2023epu, NA49:2002pzu} yields.
The suppression of the $K^*(892)^0$ production (as expected due to rescattering processes) when going from $p$+$p$ to nucleus-nucleus collisions is observed for the two highest collision energies. In contrast, there is no such suppression for the lowest energy. The $K^*/K$ ratio in Ar+Sc at $\sqrt{s_\mathrm{NN}}$ $\approx$ 17 GeV is similar to the results from Pb+Pb collisions.

\begin{figure}[!ht]
    \centering
        \includegraphics[width=0.49\linewidth]{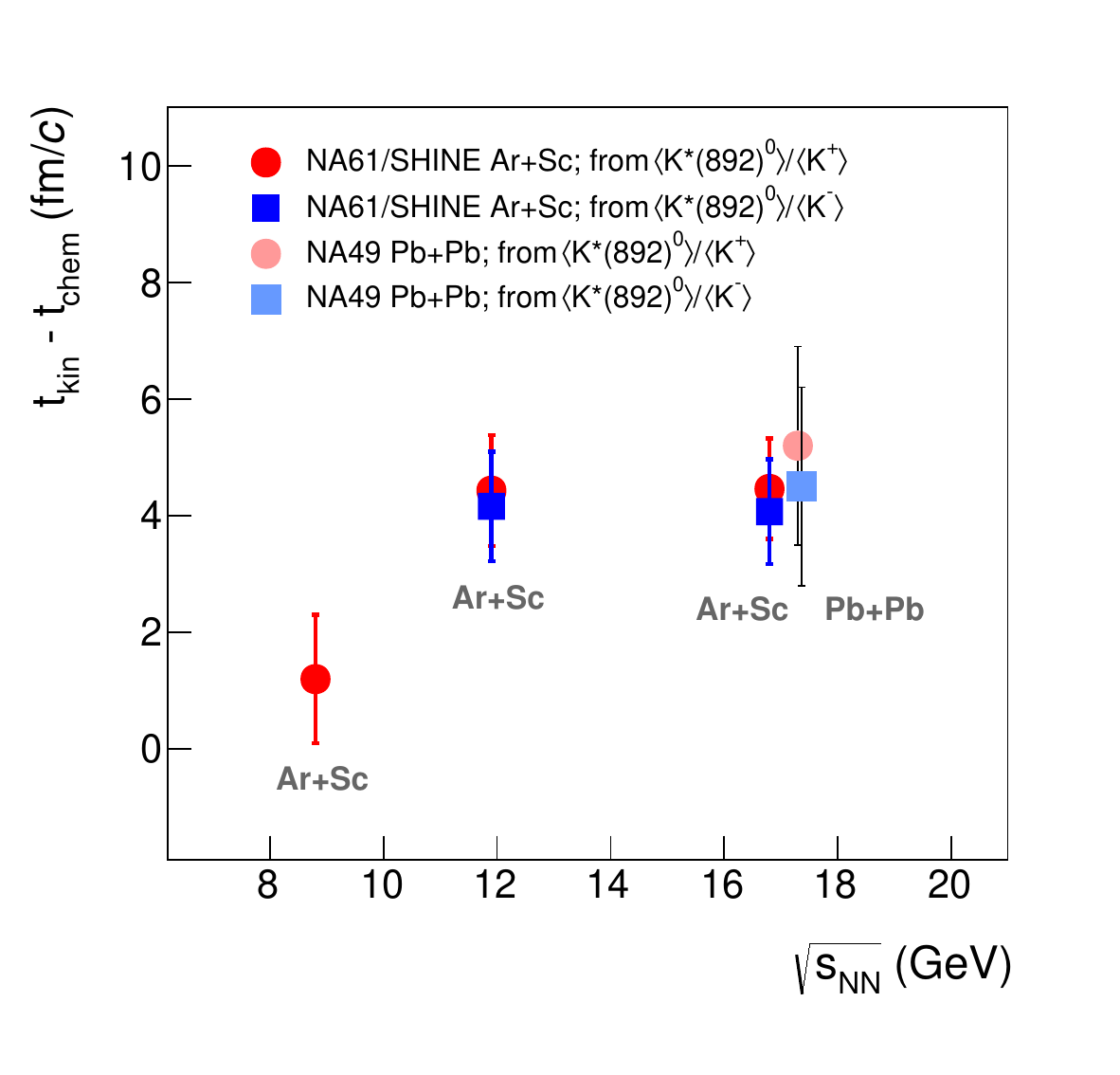}
 \vspace{-0.5cm}
    \caption{Time between freeze-outs as a function of collision energy for central Ar+Sc and Pb+Pb collisions. Vertical bars represent total uncertainties.}
    \label{fig:time}
\end{figure}

The $K^*/K$ ratio obtained from $p$+$p$ and nucleus-nucleus collisions can be used to estimate the time between freeze-outs in nucleus-nucleus collisions. Assuming no regeneration processes, the time between freeze-outs can be determined using formula~\cite{Markert:2002rw,STAR:2004bgh}:
\begin{equation}
\dfrac{K^*}{K}|_{\mathrm{kinetic}} = \dfrac{K^*}{K}|_{\mathrm{chemical}} \cdot e^{-\frac{\Delta t}{\tau}},
\end{equation}
where $\dfrac{K^*}{K}|_{\mathrm{chemical}}$ is the $\dfrac{K^*}{K}$ ratio in $p$+$p$ collisions,
$\dfrac{K^*}{K}|_{\mathrm{kinetic}}$ is the $\dfrac{K^*}{K}$ ratio in central nucleus-nucleus collisions,
$\tau$ = 4.17 fm/$c$ is the $K^*(892)^0$ mean lifetime,
and $\Delta t$ is the time between freeze-outs (in the $K^*$ rest frame).
Figure~\ref{fig:time} shows the estimated time between freeze-outs as a function of collision energy. The vertical axis presents the $\Delta t$ values boosted by the Lorentz factor:
\begin{equation}
\gamma = \sqrt{1+(\langle p_\mathrm{T} \rangle / m_0 c)^2}
\end{equation}
as suggested in Ref.~\cite{ALICE:2019xyr}. The $\langle p_\mathrm{T} \rangle$ represents the $K^{*}(892)^0$ mean transverse momentum and $m_0$ is the $K^{*}(892)^0$ rest mass.  

\section{Summary}
The preliminary results on $K^*(892)^0$ production in 0--10\% central Ar+Sc collisions at beam momenta 40$A$, 75$A$, and 150$A$ GeV/$c$ were presented. Transverse momentum, transverse mass, and rapidity spectra were obtained. The measured rapidity spectra cannot be described by the EPOS and FTFP-BERT models. The obtained $K^*/K$ ratios show an expected suppression of $K^*(892)^0$ production at beam momenta 75$A$ and 150$A$ GeV/$c$, whereas at 40$A$ GeV/$c$ no suppression is observed. The estimated times between freeze-outs in Ar+Sc collisions at beam momenta 75$A$ and 150$A$ GeV/$c$ are similar and, at the higher beam momentum, are close to the results from Pb+Pb collisions.

\vspace{0.2cm}
\noindent
\begin{footnotesize}
This work was supported by the Polish Ministry of Science and Higher Education (WUT ID-UB) and the Polish Minister of Education and Science (contract No. 2021/WK/10).
\end{footnotesize}

\end{document}